\documentstyle[prb,aps,preprint]{revtex}
\title{Aspects of the normal state phase of copper oxide  planes in high Tc superconductors.}
\author{B.C. den Hertog and M.P. Das   \\
\small{{\it Department of Theoretical Physics,}} \\
\small{{\it  Research School of Physical 
Sciences and Engineering,}} \\
\small{{\it The Australian National University, 0200  Australia.}}}

\newcommand{\sss}{\scriptscriptstyle}
\newcommand{\cid}{c^{{\scriptscriptstyle \dag}}_{i\sigma}}
\newcommand{\cad}{c^{{\scriptscriptstyle \dag}}_{{\scriptscriptstyle A}\sigma}}
\newcommand{\cbd}{c^{{\scriptscriptstyle \dag}}_{{\scriptscriptstyle B}\sigma}}

\newcommand{\cxo}{c^{}_{x_{1}{\bf k}\sigma}}
\newcommand{\cxt}{c^{}_{x_{2}{\bf k}\sigma}}
\newcommand{\cyo}{c^{}_{y_{1}{\bf k}\sigma}}
\newcommand{\cyt}{c^{}_{y_{2}{\bf k}\sigma}}
\newcommand{\ci}{c^{}_{i\sigma}}
\newcommand{\cak}{c^{{\scriptscriptstyle \dag}}_{{\scriptscriptstyle A}{\bf k}\sigma}}
\newcommand{\cbk}{c^{{\scriptscriptstyle \dag}}_{{\scriptscriptstyle B}{\bf k}\sigma}}

\newcommand{\cjd}{c^{{\scriptscriptstyle \dag}}_{j\sigma}}
\newcommand{\cj}{c^{}_{j\sigma}}
\newcommand{\ea}{e_{{\scriptscriptstyle A}}}
\newcommand{\eb}{e_{{\scriptscriptstyle B}}}
\newcommand{\su}{s_{{\scriptscriptstyle \uparrow}}}
\newcommand{\sd}{s_{{\scriptscriptstyle \downarrow}}}

\newcommand{\da}{d_{{\scriptscriptstyle A}}}
\newcommand{\db}{d_{{\scriptscriptstyle B}}}

\newcommand{\ls}{(\frac{1}{2}-u_{j})}
\newcommand{\lx}{(\frac{1}{2}+u_{j})}
\newcommand{\sma}{{\scriptscriptstyle A}}
\newcommand{\smb}{{\scriptscriptstyle B}}
\begin{document}
\maketitle
\vspace*{1cm}
PACS Numbers: 74.25.Jb, 74.25.Kc,  71.27.+a
\begin{abstract}
	We examine various aspects of the normal state phase of 
         ${\rm CuO_{2}}$ planes
	 in the high Tc superconductors. In particular, within the context of 
	 the three band  Hubbard model, we study as a function of 
	 doping the competition between a charge density wave phase induced 
	 by oxygen breathing modes, antiferromagnetic order and 
	 paramagnetism. To account for the strong electronic interactions, we 
	 use the finite $U$ slave boson method of Kotliar and Ruckenstein.
\end{abstract}
\section{Introduction}
Since the discovery of 
 high temperature superconductivity in the cuprates, there has  been 
 a tremendous effort to deduce the properties of  strongly interacting 
 fermion systems in two-dimensions (2D). Regarding the copper oxide planes, 
 this has led to a strong focus on purely electronic versions of the one and 
 three band Hubbard models and their variants. Whilst this has led to 
 progress in understanding the insulating behaviour of the high Tc materials 
 at half filling (as charge transfer insulators in the three band 
 case and as Mott-Hubbard  insulators in the one band case), there 
 is little consensus as to whether superconductivity  exists in 
  either of these models or whether they are  Fermi liquids or something more 
 exotic. Furthermore, there have been  relatively few attempts to
  understand the role of the   underlying lattice  in  
these systems. The motivation to study this last point becomes more acute with 
the realization that  there is strong experimental evidence to suggest
that  a 
coupling does exist between  charge carriers and the lattice in  
these materials  
\cite{Arai,Friedl,Krantz,Litvinchuk,Bruesch,Steiner}.

 From a theoretical viewpoint there have been several studies of both  the one  and three band
 Peierls Hubbard models in 2D. Here, we present a brief account of 
 some recent results  in this area. Dobry et al. \cite{Dobry} 
 suggest using  exact diagonalisation of  a twelve site ${\rm 
 Cu_{4}O_{8}}$ 
 cluster, that in the stoichiometric state (zero doping), 
 for any reasonable coupling of the fermions to the 
 in-plane oxygen breathing modes and in the adiabatic limit, 
 the system is stable against a lattice deformation. 
 They also show that doping with one hole can produce an extended charge 
 density wave (CDW) below a critical coupling to the lattice, and a self 
 trapped polaron which forms a more generalised Zhang-Rice singlet 
 above the critical coupling. On the other hand, 
in their Hartree Fock analysis of 
 a system  of 6x6 ${\rm CuO_{2}}$ unit cells,  Yonemitsu et al. 
 \cite{yonemitsu1,yonemitsu2} show  that in the zero doped state the 
 ${\rm CuO_{2}}$ planes can, above a critical coupling strength, 
 undergo some type of lattice deformation, ie bond 
 order wave, CDW. Slave boson studies of the one band 
 Peierls Hubbard model \cite{Feshke,Deeg} show that at half filling a 
 stable paramagnetic phase with an on-site frozen breathing mode induced 
 long range CDW occurs below a critical value of the 
 Hubbard repulsion $U$. Mitra and Behera  show \cite{Mitra} that in general 
 electron correlation suppresses the CDW state.

The purpose of this paper is to  examine systematically as a function 
of doping
the stability of the three band   Hubbard 
model \cite{Emerya,Littlewood,Varma,Emeryb} against a lattice deformation 
created by the in-plane oxygen breathing 
modes. More to the point, we  wish to study whether this model is susceptible to at least a paramagnetic 
CDW phase either at zero or finite doping, and more 
importantly, whether  the copper oxide planes 
can  develop from a homogenous antiferromagnetic phase in the undoped system 
to a paramagnetic CDW phase at finite doping.

Using the slave boson method of Kotliar and Ruckenstein (KR) \cite{Kotliar},
 we are able to take to into account the strong on-site and nearest 
 neighbour fermion interactions. 
This finite Hubbard $U$ technique has the 
advantage of being able to cope with the strong coupling regime in a 
non-perturbative way, and at 
the same time it is in reasonable agreement with Monte Carlo results 
\cite{Lilly} even at the mean field level. Thus, armed with this apparatus, we 
are able to examine as a function of doping and interaction 
parameters at the mean field level, the competition between an undistorted 
antiferromagnetic state, the normal paramagnetic state and a 
paramagnetic CDW state induced by lattice deformations. Lastly, because we use the  three 
band model,  we are able to accommodate a more  complete tight binding 
band structure which includes oxygen-oxygen hopping  in order 
to produce a realistic Fermi surface \cite{Sancho}.

\section{Model Calculations}

To examine the possibility of a  lattice instability, we include  
 an electron-phonon coupling which modifies 
the copper-oxygen hopping strength depending on whether an oxygen ion 
is displaced towards a copper ion or  away from it 
\cite{Dobry,yonemitsu1,yonemitsu2}. 
To simplify calculations we make the following approximations. 
Firstly, we  assume a frozen phonon picture for  the oxygen ions and thus 
work in the adiabatic limit. Secondly, we assume that the 
oxygen-oxygen hopping is not modified by the  relative displacements of the oxygen ions.
 With these provisions, the three 
band Hubbard Hamiltonian with electron-phonon coupling (three 
band Peierls Hubbard model) is given by
\begin{eqnarray}
 \label{hamil}
 H&=&\sum_{\langle ij\rangle\sigma} t^{}_{ij} 
 (c^{{\scriptscriptstyle \dag}}_{i\sigma}c^{}_{j\sigma} + h.c.) 
 +\sum_{\langle jj^{\prime}\rangle\sigma}
 t^{}_{jj^{\prime}}(c^{{\scriptscriptstyle \dag}}_{j\sigma}c^{}_{j'\sigma} + h.c.)
 + \sum_{i\sigma}\epsilon^{}_{i}\cid\ci  
 +\sum_{j\sigma}\epsilon^{}_{j}\cjd\cj   \nonumber \\
 &+&  \sum_{i\sigma}U n_{i\uparrow}n_{i\downarrow} + 
 \sum_{\langle ij\rangle\sigma\sigma^{\prime}}V n_{i\sigma}n_{j\sigma^{\prime}} 
 +\sum_{j}\frac{1}{2} K u_{j}^{2} \;\; , 
\end{eqnarray}
where  $\cid$ creates a hole at a copper site $i$ with spin $\sigma$ and
 $\cjd$ creates a hole at an oxygen site $j$ with spin $\sigma$.
 The on-site copper (oxygen) energy is given 
by $\epsilon_{i} \; (\epsilon_{j})$. The hopping between copper and 
oxygen sites is represented by $t_{ij}$ and its sign depends on the 
symmetry of the copper $d_{x^{2}-y^{2}}$ and oxygen $p_{x}$ and 
$p_{y}$ orbitals. The sum over $\langle ij\rangle$ refers to  nearest neighbour
 copper and oxygen sites only. Without lattice deformation 
and in the nearest neighbour hopping picture, $t_{ij}$ would be given 
by a constant $t$. 
However, $t_{ij}$ is assumed to be modified by the in-plane breathing modes  as stated above in the following manner 
\cite{yonemitsu1,yonemitsu2}. If the surrounding oxygen 
sites are deformed towards a copper site, then $t_{ij}=t + 
\alpha u_{j}$, where $u_{j}$ is the absolute value of the oxygen 
lattice displacement from its mean position. Similarly, if the 
surrounding oxygens are displaced away from a copper site, then the 
hopping becomes $t_{ij}=t-\alpha u_{j}$. The parameter 
$t_{jj^{\prime}}$  in (\ref{hamil})
represents oxygen-oxygen hopping and its sign also depends on the symmetry 
of the oxygen orbitals. The sum over $\langle jj^{\prime}\rangle$ refers to 
hopping between pairs of second nearest neighbour 
oxygen $j$ and  $j^{\prime}$ sites. The $U$ term 
represents the Hubbard interaction at copper sites whilst $V$ is the 
nearest neighbour interaction between holes at 
adjacent copper and oxygen sites. We 
neglect an on-site interaction at the oxygen sites since the 
probability of double occupancy  of these  sites by holes for physically 
reasonable hopping and interaction parameters and not too large doping is quite small. 
The classical force constant of the oxygen ions is represented by 
the parameter $K$. Following references
\cite{Dobry,yonemitsu1,yonemitsu2} we introduce a dimensionless 
electron-phonon coupling constant $\lambda\equiv\alpha^{2}/Kt$. 

To accommodate the strong on-site and nearest neighbour fermion 
interactions, we introduce four auxiliary  bosons fields
\cite{Kotliar} for the copper sites. In this scheme, each possible 
electronic configuration of the copper site is represented by one of
these fields. The  fermionic creation operator is mapped 
$\cid\mapsto q^{}_{i\sigma}\cid$ where $q^{}_{i\sigma}$ is defined as
\begin{equation}
q^{}_{i\sigma}\equiv\frac{e^{}_{i}s^{{\sss \dag}}_{i\sigma} + s^{}_{i\bar{\sigma}} 
d^{{\sss \dag}}_{i}}{\sqrt{(1-d^{{\sss \dag}}_{i}d^{}_{i} - 
s^{{\sss \dag}}_{i\sigma}s^{}_{i\sigma})(1 - e^{{\sss \dag}}_{i}e^{}_{i} - 
s^{{\sss \dag}}_{i\bar{\sigma}}s^{}_{i\bar{\sigma}})}} \;\; . 
\end{equation}
The boson fields $e_{i},s_{i\sigma},$ and $d_{i}$ represent empty, 
single with spin $\sigma$ and double occupation of copper sites  respectively.
 The square root terms in $q_{i\sigma}$ ensures that the mapping becomes 
trivial at the mean field level in the non-interacting limit 
$\;U\rightarrow 0,V \rightarrow 0$. 
The KR slave boson method 
replaces the Hubbard interaction  with a bilinear term involving 
the boson double occupancy fields whilst renormalising the hopping 
energy of the fermion quasiparticles by the factor $q_{i\sigma}$. 
Unphysical states arising from the enlargened Hilbert space are 
eliminated via the following constraints;
\begin{equation}
\cid\ci=s^{{\sss \dag}}_{i\sigma}s^{}_{i\sigma} + d^{{\sss \dag}}_{i}d^{}_{i}
\end{equation}
and
\begin{equation}
e^{{\sss \dag}}_{i}e^{}_{i} + \sum_{\sigma}s^{{\sss \dag}}_{i\sigma}s^{}_{i\sigma} +
 d^{{\sss \dag}}_{i}d^{}_{i} - 1 =0 \; \; ,
\end{equation}
which represent charge conservation and the completeness of the 
bosonic operators respectively.
With this mapping in place the   Hamiltonian becomes;
\begin{eqnarray}
H&=&\sum_{\langle ij\rangle\sigma}t^{}_{ij}(q^{}_{i\sigma}\cid\cj + h.c.) 
+\sum_{\langle jj^{\prime}\rangle\sigma} 
t^{}_{jj^{\prime}}(c^{{\sss \dag}}_{j\sigma}c^{}_{j^{\prime}\sigma}+ h.c.)  
+ \sum_{i\sigma}(\epsilon^{}_{i}+ \gamma^{}_{i\sigma})\cid\ci \nonumber \\
&+& \sum_{j\sigma}\epsilon^{}_{j}\cjd\cj 
+\sum_{i}U d^{{\sss \dag}}_{i}d^{}_{i}
+ \sum_{\langle ij\rangle\sigma\sigma^{\prime}}
 V (s^{{\sss \dag}}_{i\sigma}s^{}_{i\sigma} + 
d^{{\sss \dag}}_{i}d^{}_{i}) 
c^{{\sss \dag}}_{j\sigma^{\prime}}c^{}_{j\sigma^{\prime}} 
+\sum_{j}\frac{1}{2} K u_{j}^{2} \nonumber \\
&-& \sum_{i\sigma}\gamma^{}_{i\sigma}(s^{{\sss \dag}}_{i\sigma}s^{}_{i\sigma} + 
d^{{\sss \dag}}_{i}d^{}_{i}) + 
\sum_{i}\gamma^{\prime}_{i}(e^{{\sss \dag}}_{i}e^{}_{i} + 
\sum_{\sigma}s^{{\sss \dag}}_{i\sigma}s^{}_{i\sigma} +
 d^{{\sss \dag}}_{i}d^{}_{i}-1) \; \; ,
\end{eqnarray}
where $\gamma_{i\sigma}$ and $\gamma^{\prime}_{i}$ are Lagrange multipliers 
enforcing the constraints on the boson fields.

The partition function for the system can be expressed as a 
functional integral over complex (bosons) and Grassmann fields (fermions). 
Integrating exactly over the Grassmann fields leaves the partition 
function as\cite{Negele}; 
\begin{equation}
Z=\int 
D[e]D[s_{\sigma}]D[d]D[\gamma]D[\gamma^{\prime}]e^{-S^{boson}_{eff}} \; \; ,
\end{equation}
where the bosonic action is
\begin{eqnarray}
 S^{boson}_{eff}&=&\int_{0}^{\beta}d\tau \;\mbox{\LARGE \{ }{\rm Tr} 
 \ln\mbox{\Large [} t^{}_{ij}q^{}_{i\sigma} + 
(\gamma^{}_{i\sigma} + \epsilon^{}_{i} + 
\partial^{}_{\tau})\delta^{}_{ji} 
\nonumber \\
&+&
 \mbox{\Large (}\epsilon^{}_{j} + \partial_{\tau} + 
V\sum_{nn\sigma^{\prime}} 
(s^{{\sss \dag}}_{nn\sigma^{\prime}}s^{}_{nn\sigma^{\prime}} 
+d^{{\sss \dag}}_{nn}d^{}_{nn})\mbox{\Large )}\delta^{}_{ij} + 
t^{}_{jj^{\prime}}\delta^{}_{ij^{\prime}}\mbox{\Large ]}
\nonumber \\
&+&\sum_{i\sigma}s^{\dag}_{i\sigma} (\gamma_{i}^{\prime} + \partial^{}_{\tau} - 
\gamma^{}_{i\sigma})s^{}_{i\sigma} + 
\sum_{i\sigma}d^{{\sss \dag}}_{i}(U  + 
\gamma_{i}^{\prime} + 
\partial^{}_{\tau} - \gamma^{}_{i\sigma})d^{}_{i} \nonumber \\
&+&  \sum_{i}e^{{\sss \dag}}_{i} 
(\gamma_{i}^{\prime} + \partial^{}_{\tau})e^{}_{i} 
 -\sum_{i}\gamma_{i}^{\prime} + 
 \sum_{j}\frac{1}{2}Ku_{j}^{2}\mbox{  \LARGE \}} \; \;  
\end{eqnarray}
and the subscript $nn$ implies nearest neighbour summations only.

To evaluate the partition function we employ the saddle point 
approximation \cite{Kotliar} where all the boson fields are $c$ 
numbers. Accordingly, at $T=0$ the energy per cell is given by
\begin{eqnarray}
\label{epercell}
E&=&\frac{1}{N}\sum_{\langle ij\rangle\sigma} 
q^{}_{i\sigma}t^{}_{ij}\langle\cid\cj + h.c.\rangle + 
\frac{1}{N}\sum_{\langle jj^{\prime}\rangle\sigma} 
t^{}_{jj^{\prime}}\langle\cjd c^{}_{j^{\prime}\sigma}+ 
h.c.\rangle \nonumber \\
&+&\frac{1}{N} \sum_{i\sigma}(\epsilon^{}_{i} + 
\gamma^{}_{i\sigma})n^{}_{i\sigma} + 
\frac{1}{N}\sum_{j\sigma}\epsilon^{}_{j}n^{}_{j\sigma} \nonumber \\
&+&\frac{1}{N}\sum_{i}U d^{2}_{i} + 
\frac{1}{N}\sum_{\langle ij\rangle\sigma\sigma^{\prime}}V(s^{2}_{i\sigma} + 
d^{2}_{i})n^{}_{j\sigma^{\prime}} + 
\frac{1}{N}\sum_{j}\frac{1}{2}Ku_{j}^{2} \nonumber \\
&-&\frac{1}{N}\sum_{i\sigma}\gamma^{}_{i\sigma}(s^{2}_{i\sigma} + 
d^{2}_{i}) + \frac{1}{N}\sum_{i}\gamma^{\prime}_{i}(e^{2}_{i} + 
\sum_{\sigma}s^{2}_{i\sigma} + d^{2}_{i} -1 ) \;\; ,
\end{eqnarray}
where $N$ is the number of unit cells.

Equation (\ref{epercell}) must  be minimized with respect to the boson 
fields. The CDW phase can be examined by dividing the 
copper lattice into two sub-lattices A and B, which results 
in a doubling of the unit cell and hence a halving of the Brillouin 
zone. The above energy per cell will 
be modified by inclusion of summations over the sub-lattices and by 
dividing the boson fields into $e_{{\sss \nu}},s_{{\sss \nu}}$ and 
$d_{{\sss \nu}}$ where $\nu=A,B$. Here we have assumed a paramagnetic CDW phase, 
thus $s_{\nu\sigma}=s_{\nu\bar{\sigma}}$. 

To  examine the purely 
antiferromagnetic state, the copper lattice is  again divided and the 
unit cell doubled. The displacement of the oxygen ions is set to zero. 
Due to the symmetry of such a phase, 
$s_{{\scriptscriptstyle A}\sigma}=s_{{\scriptscriptstyle B}\bar{\sigma}}$, 
$\gamma_{{\scriptscriptstyle A}\sigma}=\gamma_{{\scriptscriptstyle B}\bar{\sigma}}$ and 
$d_{{\scriptscriptstyle A}}=d_{{\scriptscriptstyle B}}$. The 
homogeneous
paramagnetic phase can be studied by a straight forward minimization of 
the above energy per cell.
\subsection{Charge Density Wave Phase.}
In this subsection we derive the self consistent minimization equations 
for the CDW phase. Dividing the copper lattice  into 
an $A$ and $B$ sub-lattices and minimizing the energy per ${\rm Cu_{2}O_{4}}$ 
cell with respect to $u_{j}$, we get
\begin{equation}
\frac{1}{M}\frac{\partial}{\partial u_{j}}
\left[ \sum_{\langle{\scriptscriptstyle A}j\rangle\sigma}q^{}_{{\scriptscriptstyle A}} 
t^{}_{{\scriptscriptstyle A}}\langle\cad\cj + h.c.\rangle + 
\sum_{\langle{\scriptscriptstyle B}j\sigma\rangle}q^{}_{{\scriptscriptstyle B}} 
t^{}_{{\scriptscriptstyle B}}\langle\cbd\cj + h.c.\rangle\right] + 
4Ku_{j}=0 \; \; ,
\end{equation}
where $M$ is the number of ${\rm Cu_{2}O_{4}}$ cells.
Moving to Bloch functions the above equation becomes;
\begin{equation}
\label{ming}
2\alpha(q_{{\scriptscriptstyle A}}\Omega_{{\scriptscriptstyle A}} - 
q_{{\scriptscriptstyle B}}\Omega_{{\scriptscriptstyle B}}) + 
q_{{\scriptscriptstyle A}}\Lambda_{{\scriptscriptstyle A}} + 
q_{{\scriptscriptstyle B}}\Lambda_{{\scriptscriptstyle B}} + 
4Ku_{j}=0 \;\; ,
\end{equation} 
where
\begin{eqnarray}
\Omega_{\sma}&=&\frac{1}{2M}\sum_{{\bf 
k}\sigma}\left (-\left[e^{ik_{x}\ls}\langle\cak\cxo\rangle + h.c.\right]
 + \left[e^{-ik_{x}\ls}\langle\cak\cxt\rangle + 
h.c.\right]\right. \nonumber \\
&+&\left.\left[e^{ik_{y}\ls}\langle\cak\cyo\rangle+ h.c.\right]
 - \left[e^{-ik_{y}\ls}\langle\cak\cyt\rangle + h.c.\right]\;\right)
\;\; ,
\end{eqnarray}
\begin{eqnarray}
\Omega_{\smb}&=&\frac{1}{2M}\sum_{{\bf 
k}\sigma}\left(\;\left[e^{-ik_{x}\lx}\langle\cbk\cxo\rangle + h.c.\right] - 
\left[e^{ik_{x}\lx}\langle\cbk\cxt\rangle + h.c.\right]\right. \nonumber \\
&-&\left.\;\left[e^{-ik_{y}\lx}\langle\cbk\cyo\rangle + h.c.\right]
 + \left[e^{ik_{y}\lx}\langle\cbk\cyt\rangle + 
h.c.\right]\;\right) \; \; ,
\end{eqnarray}
\begin{eqnarray}
\Lambda_{\sma}&=&\frac{t_{\sma}}{M}\sum_{{\bf k}\sigma}
\left(-\left[-ik_{x}e^{ik_{x}\ls}\langle\cak\cxo\rangle + h.c.\right] + 
\left[ik_{x}e^{-ik_{x}\ls}\langle\cak\cxt\rangle + h.c.\right]\right. \nonumber \\
&+& \left.\left[-ik_{y}e^{ik_{y}\ls}\langle\cak\cyo\rangle + h.c.\right] - 
\left[ik_{y}e^{-ik_{y}\ls}\langle\cak\cyt\rangle + h.c.\right ]\;\right) 
\end{eqnarray}
and
\begin{eqnarray}
\Lambda_{\smb}&=&\frac{t_{\smb}}{M}\sum_{{\bf k}\sigma} 
\left(\;\left[-ik_{x}e^{-ik_{x}\lx}\langle\cbk\cxo\rangle + h.c.\right] - 
\left[ik_{x}e^{ik_{x}\lx}\langle\cbk\cxt\rangle + h.c \right]\right. \nonumber \\
&-&\left.\left[-ik_{y}e^{-ik_{y}\lx}\langle\cbk\cyo\rangle + h.c.\right] + 
\left[ik_{y}e^{ik_{y}\lx}\langle\cbk\cyt\rangle + h.c.\right]\;\right) \; \; .
\end{eqnarray}
The hopping parameters are defined by $t_{\sma}=t + \alpha u_{j}$ 
and $t_{\smb}=t - \alpha u_{j}$. That is to say that the oxygen ions 
are breathed in towards the copper $A$ sites thus strengthening the 
hopping between them, and breathed out from the copper $B$ sites weakening 
the hopping between $B$ sites and oxygen sites. The lattice 
displacement variable is in units of the Cu-Cu lattice spacing. The 
oxygen sites are labeled around a copper $A$ site in the following way. Looking down upon the ${\rm CuO_{2}}$ plane, $x_{1}$ is the oxygen site to the right of the $A$ site, 
$x_{2}$ is to the left, $y_{1}$ is above and $y_{2}$ is below.

 Minimizing equation (\ref{epercell}) with respect to the boson fields 
 on the $A$ and $B$ sub-lattices, we obtain the following equations;
\begin{equation}
t^{}_{\nu}\Omega^{}_{\nu}\frac{\partial q_{\nu}}{\partial d_{\nu}} + 
Ud^{}_{\nu} + 8V n^{}_{j}d^{}_{\nu} - 2\gamma^{}_{\nu}d^{}_{\nu} + 
d^{}_{\nu}\gamma_{\nu}^{\prime}=0 \; \; ,
\end{equation}
\begin{equation}
t^{}_{\nu}\Omega^{}_{\nu}\frac{\partial q_{\nu}}{\partial s_{\nu}} + 
8Vn^{}_{j}s^{}_{\nu} -2 \gamma^{}_{\nu}s^{}_{\nu} + 2 \gamma_{\nu}^{\prime} 
s^{}_{\nu} =0 \; \; 
\end{equation}
and
\begin{equation}
t^{}_{\nu}\Omega^{}_{\nu}\frac{\partial q_{\nu}}{\partial e_{\nu}} + 
\gamma_{\nu}^{\prime} e^{}_{\nu} =0 \; \; ,
\end{equation}
where $\nu= A,B$. Combining these equations with the constraints we 
arrive at the minimization equations for the $A$ and $B$ sub-lattices;
\begin{equation}
\label{mind}
U=\frac{t_{\nu}\Omega_{\nu}}{\sqrt{\frac{n_{\nu}}{2}(1 - 
\frac{n_{\nu}}{2})}}\left[\frac{1 + 2d_{\nu}^{2} - \frac{3 
n_{\nu}}{2}}{\sqrt{(1 + d_{\nu}^{2} - n^{}_{\nu}) 
(\frac{n_{\nu}}{2}-d_{\nu}^{2})}} + \frac{2d_{\nu}^{2} - 
\frac{n_{\nu}}{2}}{\sqrt{d_{\nu}^{2}(\frac{n_{\nu}}{2} - 
d_{\nu}^{2})}}\right] \; \; .
\end{equation}
The total energy per cell is thus minimized by self consistently 
diagonalizing the fermion part of the effective Hamiltonian in order 
to calculate $n^{}_{\nu}, \Omega^{}_{\nu},\Lambda^{}_{\nu},$ and $u_{j}$ and solving equations 
(\ref{ming}) and (\ref{mind}). 

In presenting results for this phase, we begin by studying the interplay 
between Coulomb interaction and 
electron-phonon coupling in the undoped state. 
We have used the   parameter set \cite{Hybertson} of $t=1.3,\; 
t_{jj^{\prime}}=0.65,\; V=1.2,\;$ and $ \epsilon_{j}-\epsilon_{i}=3.6$ 
where all energies are in eV. The force constant $K=32 t$ 
eV/\AA $\;\;$ \cite{yonemitsu1}. This 
parameter set is used for all calculations. Of course the overall stability of 
the paramagnetic CDW phase can  only be considered after 
comparing the total energy to that of the homogenous paramagnetic and 
antiferromagnetic phases, which we will consider in later sections. 
 
 Fig. 1 shows the charge density difference between 
copper $A$ and $B$ sites as a function of $U$ for different electron 
phonon coupling strengths. In this CDW phase, the electron 
phonon coupling has increased the copper $A$ site charge density due 
to the increased hopping amplitude $t_{\sma}$. Conversely, the charge 
density at the $B$ sites has been decreased significantly due to 
$t_{\smb}$. This process is hindered by the Hubbard repulsion at the 
copper sites, which naturally works against this increased charge 
density at the $A$ sites. 
The results of Fig. 1 seem to indicate that at  a 
certain value of $U$, for a given $\lambda$, the system undergoes a first order transition to 
the homogenous paramagnetic state. At this point the Coulomb repulsion 
has become too strong to sustain the CDW.
As $U$ is increased form zero the $A$ site charge density decreases, 
however the $B$ site charge density does not increase at the same 
rate. There is an overall shift in charge to the oxygen sites as $U$ 
is increased. This increase is shown in Fig. 2. The charge 
continues to build up on the oxygen sites until just before 
$U_{crit}$, where there is a small reversal in charge flow. 
It can also be seen in Fig. 1 that the stability of the system as a function of $U$ is quite 
sensitive to the electron-phonon coupling strength. 
 Fig. 1  shows that for a reasonably strong coupling $(\lambda>0.8)$ 
the system is quite robust (at least there is a local minimum in the 
energy) for any physically reasonable value of $U$ for the ${\rm CuO_{2}}$
 planes. 
 
 In Fig. 3 we show the difference in the charge densities on copper $A$ 
 and $B$ sites $\Delta\;(\equiv n^{}_{\sma} - n^{}_{\smb})$ as a function of
  doping $(\delta\equiv n-1)$ for the parameter sets
  $\lambda=1.00$, $U=10$ eV  and $\lambda=0.95$, $U=6$ eV. There is an almost linear decrease 
 in the CDW order parameter as the system is doped until one reaches 
 $\delta_{crit}\approx 0.49$ for both parameter sets. Prior to $\delta^{}_{crit}$, the $B$ site 
 charge is almost localised due to an extremely small effective 
 hopping $q^{}_{\smb}t_{\smb}\sim 0.1$. The double occupancy on 
 the $B$ sites is such that $d_{\smb}^{2}\sim 10^{-4}$, and remains 
 at this magnitude  as more charge builds up on the $B$ sites as the system is 
 doped. Because of this, the holes on the $B$ sites become more and 
 more localised as extra holes are added to the systems since 
 $q^{}_{\smb}\rightarrow 0$ as $n^{}_{\smb}\rightarrow 1$ and 
 $d^{}_{\smb}\rightarrow 0 \;$ \cite{den Hertog}. As 
 $\delta\rightarrow\delta^{}_{crit}$, $n^{}_{\smb}\rightarrow 1$, thus 
 when $\delta$ exceeds $ \delta^{}_{crit}$ and therefore $n^{}_{\smb}>1$ 
 the system is required to obey the physical constraint 
 $d^{2}_{\smb}>n^{}_{\smb}-1$, causing a finite Hubbard energy on the 
 $B$ sites which makes the system in this phase unstable.

 \subsection{Antiferromagnetic Phase.}
 There have been previous slave boson
 calculations carried out on the antiferromagnetic 
 and paramagnetic phases of the three band Hubbard model by Zhang et al
  \cite{Zhang}.
 However, they have used a parameter set  containing a small 
 oxygen-oxygen hopping parameter ($t_{jj^{\prime}}=0.2 $ eV) and charge 
 transfer energy ($\epsilon_{j}-\epsilon_{i}=1.5$ eV). Also, their 
 value of $U$ ($6$ eV), is just at the lower bound of what is a reasonable 
 value for the Hubbard interaction on copper sites in the
 copper oxide planes \cite{Mila}. This being the case, it is useful to 
 redo and expand on some of these types of  calculations 
 with improved estimates of the hopping and interaction parameters, 
 especially when comparing the energies of these phases  to a phonon 
 driven one.

 The saddle point equations in the antiferromagnetic phase are 
 derived by again dividing the copper lattice into $A$ and $B$ 
 sub-lattices. As stated before, the symmetry of this phase implies 
 $\ea=\eb, s_{{\scriptscriptstyle A\uparrow}}=s_{{\scriptscriptstyle 
 B\downarrow}}\equiv\su, s_{{\scriptscriptstyle A\downarrow}}= 
 s_{{\scriptscriptstyle B\uparrow}}\equiv\sd$ and $\da=\db\equiv d$. 
 Following a procedure similar to that in the CDW 
 section, the following minimization equation is derived;
 \begin{eqnarray}
 \label{minaf}
 U&=&\frac{\Omega_{{\sss \uparrow}}t}{\sqrt{(1-n_{i{\sss \uparrow}})n_{i{\sss \uparrow}}}}
 \left[\frac{1 + 2d^{2}-n_{i}-n_{i{\sss \uparrow}}}{\sqrt{(1 + 
 d^{2}-n_{i})(n_{i{\sss \uparrow}}-d^{2})}} + 
 \frac{2d^{2}-n_{i{\sss \downarrow}}}{\sqrt{d^{2}(n_{i{\sss \downarrow}}-d^{2})}}\right] \nonumber \\
 &+&\frac{\Omega_{{\sss \downarrow}}t}{\sqrt{(1-n_{i{\sss \downarrow}})n_{i{\sss \downarrow}}}} 
 \left[ \frac{1+2d^{2}-n_{i}-n_{i{\sss \downarrow}}}{\sqrt{ 
 (1+d^{2}-n_{i})(n_{i{\sss \downarrow}}-d^{2})}} + 
 \frac{2d^{2}-n_{i{\sss \uparrow}}}{\sqrt{d^{2}(n_{i{\sss \uparrow}}-d^{2})}}\right]
\; \; ,
\end{eqnarray}
where 
\[
\Omega_{{\sss \uparrow}}\equiv
\frac{1}{2M}\sum_{\langle\sma j\rangle} 
(\langle c^{{\sss \dag}}_{\sma{\sss \uparrow}}c^{}_{j{\sss \uparrow}}\rangle + h.c.)
=\frac{1}{2M}\sum_{\langle\smb j\rangle} 
(\langle c^{{\sss \dag}}_{\smb{\sss \downarrow}}c^{}_{j{\sss \downarrow}}\rangle
 + h.c.) \;\; ,
\]
\[
\Omega_{{\sss \downarrow}}\equiv
\frac{1}{2M}\sum_{\langle\sma j\rangle} 
(\langle c^{{\sss \dag}}_{\sma{\sss \downarrow}}c^{}_{j{\sss \downarrow}}\rangle 
+ h.c.) 
=\frac{1}{2M}\sum_{\langle\smb j\rangle} 
(\langle c^{{\sss \dag}}_{\smb{\sss \uparrow}}c^{}_{j{\sss \uparrow}}\rangle + 
h.c.) \; \; ,
\]
and $n^{}_{i}\equiv n^{}_{\sma}=n^{}_{\smb},n^{}_{i{\sss \uparrow}}\equiv 
n^{}_{\sma{\sss \uparrow}}=n^{}_{\smb{\sss \downarrow}}$ and 
$n^{}_{i{\sss \downarrow}}\equiv n^{}_{\sma{\sss \downarrow}}=n^{}_{\smb{\sss \uparrow}}$. 
Equation (\ref{minaf}) is solved self consistently with the 
diagonalized one body fermion Hamiltonian in order to minimize the energy per 
magnetic cell. 

Fig. 4 shows the quasiparticle band structure in the 
antiferromagnetic phase at zero doping using the reference parameter set and 
$U=10$ eV. The coordinates in the figure represent the non-magnetic Brillioun zone. One can clearly see the lower and upper Hubbard bands and the 
dispersion of the oxygen bands in between. The lower Hubbard band is fully occupied 
at this hole concentration and consists of mainly 
($73\%$) of copper states. Doped holes go mainly to the oxygen sites 
and occupy states around  the $(\pi /2,\pi /2)$ region ( the magnetic 
Brillouin zone). 
It is interesting to compare this figure 
with the Hartree Fock band structure obtained by Yonemitsu  et 
al. \cite{yonemitsu2}. Although their parameter set  is 
slightly different (in units of eV; $t=1, t_{jj^{\prime}}=0.5, U=8, V=1$ and 
$U_{oxygen}=3$), the four lowest bands are similar with the same 
curvature and band minima. This is despite the fact that they include 
a Hubbard repulsion on the oxygen sites.
  For the same parameter set Fig. 5 
shows the quasiparticle band structure for a doping of $\delta=0.25$. It 
can be seen from this figure
that new states have formed within the original insulating gap of the 
undoped system reducing the gap significantly.

In Fig. 6 we show the dependence of the staggered 
magnetic moment $m \;(\equiv n^{}_{\sma{\sss \uparrow}}- n^{}_{\sma{\sss \downarrow}})$ on $U$ in the 
undoped state. As $U$ decreases 
to $\approx 1.75$ eV the lower and upper Hubbard bands begin to overlap. 
Beyond this point our results become inaccurate as the doping can not 
be pinned at zero. However, as $U$ increases from this point there is 
a steady increase in the staggered moment until it saturates for 
large $U$ at $m\approx 0.75$. Similar qualitative behaviour is found in 
the one band Hubbard model using Monte Carlo techniques \cite{White}. 
As shown by Zhang {\it et al.} \cite{Zhang} the effect of 
$t_{jj^{\prime}_{}}$ is to decrease  the magnetic moment  because 
it becomes energetically favourable for some of the holes to move on the oxygen network 
rather than occupy mainly copper sites.   
The behaviour of the staggered moment upon doping is shown in Fig. 7 
for the reference parameter set. The dotted line represents  $U=10$ eV 
and the solid line is $U=6$ eV. The local Cu moment vanishes almost 
discontinuously at $\delta_{crit}$ indicating the likelihood of a first order 
phase transition. The precise point at which the transition occurs is unclear due to the extremely slow convergence near this region. It is worth noting  that  apparently 
 $U$ as little effect on the value of $\delta$ at which the 
antiferromagnetic order vanishes. The variation of other parameters 
such as hopping and charge transfer energies have a much more 
significant effect \cite{Zhang}. The relatively small effect of varying $U$ may be 
due to the fact that at least in the undoped state, the 
antiferromagnetic order parameter has already begun to saturate at $U=6$ eV. 
\subsection{Energy per ${\rm CuO_{2}}$ cell.}
In Figs. 8 and 9 we show the calculated energy per ${\rm CuO_{2}}$ cell 
as a function of doping for the CDW, antiferromagnetic 
and paramagnetic phases. The various energies in Fig. 8 were 
calculated with an electron-phonon coupling of 
$\lambda=1.00$ and $U=10$ eV. The figure clearly shows that for this 
parameter set the system is stable against a breathing mode induced 
paramagnetic CDW 
 ground state and that as a function of doping the system 
moves from an antiferromagnetic state to a homogenous paramagnetic 
one. At this value of $\lambda$ the effective hopping parameter 
$t_{\smb}=0.07$. A  stronger coupling drives $t_{\smb}\rightarrow 0$ 
causing the copper $B$ sites to be completely localised and thus it 
becomes debatable whether it is useful to do calculations in such a 
region. However, if one believes that one can and assumes that the energy 
dependence on doping for the CDW phase
has the same gradient as shown in Fig. 8 then it may be possible to 
find a stable CDW at $\delta_{crit}>0.4$.
 The same type of calculations are shown in Fig. 9 but in this 
case $U=6$ eV and $\lambda=.95$. It is interesting to note that here the 
system starts in the antiferromagnetic phase  at zero doping then moves 
to a paramagnetic phase at $\delta\approx 0.33$ and then to a 
paramagnetic CDW phase at $\delta \approx 0.4$. These calculations 
indicate that at least in the highly doped systems, the three band 
Hubbard model is  susceptible  at this mean field level to  some sort of lattice 
distortion  (we have shown here the paramagnetic CDW) given the 
appropriate interaction parameters. However we have shown that for a larger value of $U$ and reasonably strong electron-phonon coupling, this model suppresses the CDW phase.

\section{Conclusions.}
Using the finite $U$ slave boson method of KR 
\cite{Kotliar} to account for the strong fermion-fermion interactions, 
we have examined for different interaction parameters the 
possibility of an instability of the three band 
 Hubbard model towards a paramagnetic CDW 
state induced by the oxygen breathing modes of the copper oxide planes. 
We have found that this model can at the mean field level, exhibit an instability towards  such a phase at 
large doping given a low value of the Hubbard repulsion on copper sites $U (\approx 6$ eV).  For larger $U$ the CDW is suppressed. Also, we 
have examined antiferromagnetic correlations at both zero and finite 
doping. We have shown that like the one band Hubbard model, the three 
band model antiferromagnetic order parameter saturates as a function of $U$ and 
the system undergoes a phase transition to the homogenous paramagnetic 
phase at 
finite doping.  A suggestion for further calculation would be to examine
antiferromagnetic correlations in the CDW phase, particularly its importance at low doping.

\section{Acknowledgements}
This work was initiated at the High T$_{{\rm c}}$ Superconductivity workshop 
in Canberra, Australia 1994. We thank D.C. Mattis, J.M. Wheatley and 
T.C. Choy for useful discussions. B.C.dH acknowledges the financial support of 
the Australian Federal Government in the form of an Australian Postgraduate Award Scholarship.

\newpage
\begin{center}
{\bf Figure Captions.}
\end{center}
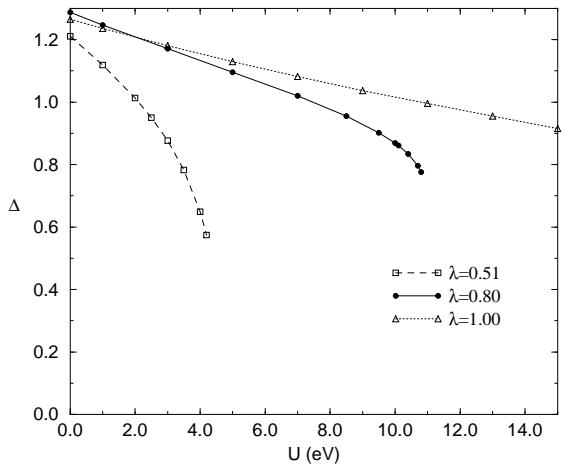
{\bf Fig 1:} Dependence of the CDW order parameter on the Hubbard 
repulsion $U$ for different electron-phonon coupling strengths and 
zero doping. \\
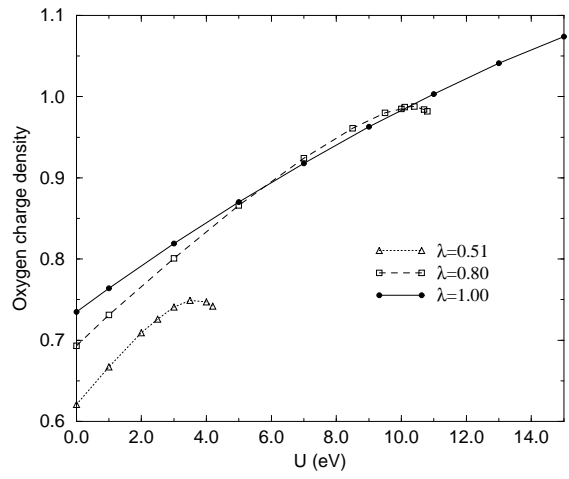
{\bf Fig 2:} Total oxygen site occupancy for a ${\rm Cu_{2}O_{4}}$ cell as a function of $U$ for different 
electron-phonon coupling strengths at zero doping. \\
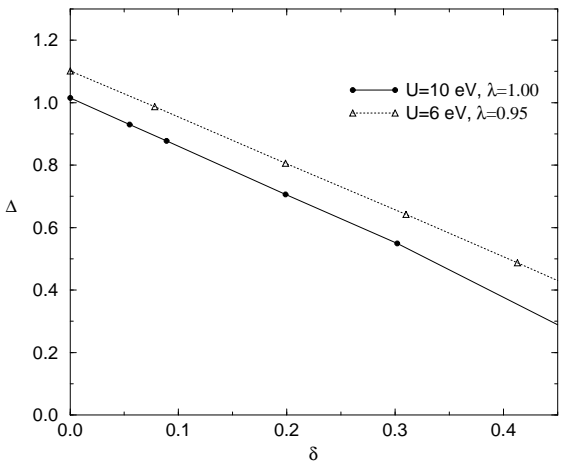
{\bf Fig 3:} Behaviour of the CDW order parameter upon doping for 
different  interaction parameter sets. \\
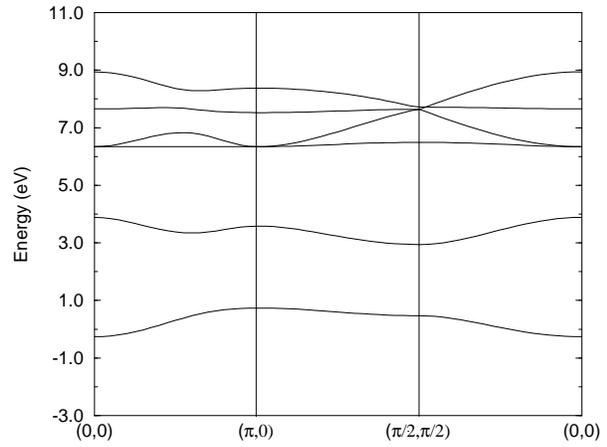
{\bf Fig 4:} Antiferromagnetic band structure for zero doping and 
$U=10$ eV. The lowest band is fully occupied whilst the remaining 
bands are empty.\\
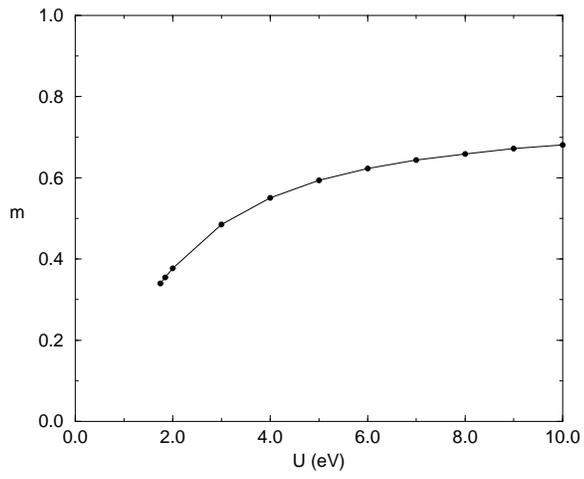
{\bf Fig 5:} Antiferromagnetic bandstructure at a doping of $\delta=0.25$ 
and $U=10$ eV. The original gap has narrowed and the second lowest 
band is partially filled. \\
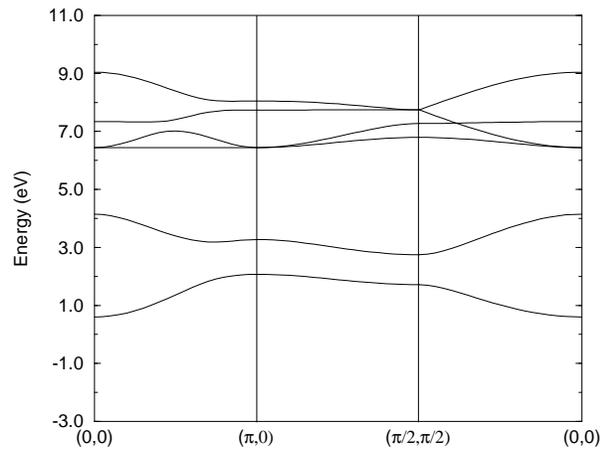
{\bf Fig 6:} Behaviour of the staggered moment on Cu sites in the 
antiferromagnetic phase as a function of $U$ at zero doping. \\
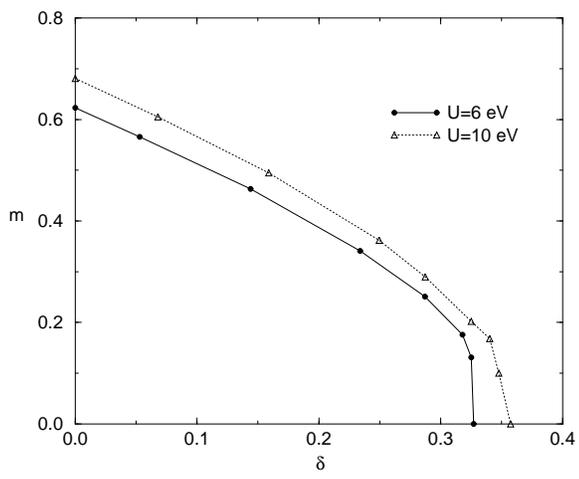
{\bf Fig 7:} Doping dependence of the staggered moment for different 
values of $U$. \\
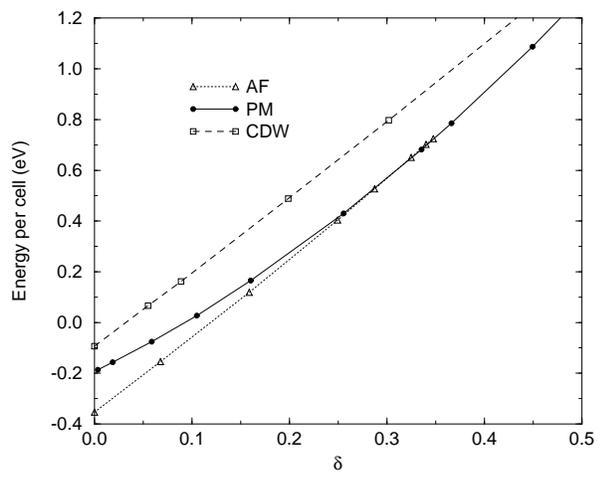
{\bf Fig 8:} Energy per ${\rm CuO_{2}}$ cell as a function of doping 
for the various phases. Here $U=10$ eV and $\lambda=1.00$. \\
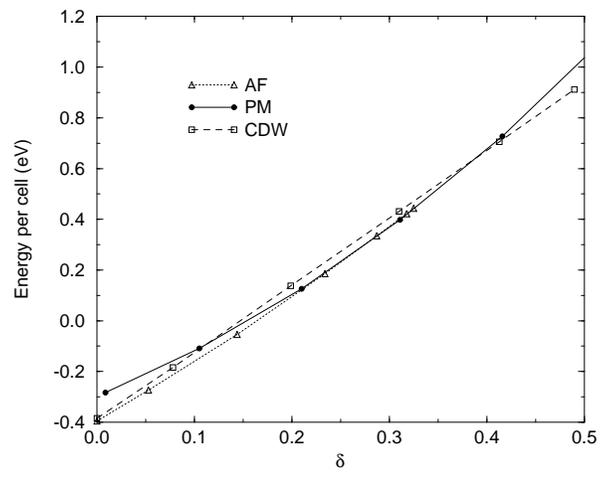
{\bf Fig 9:} Same as Fig. 8 but here the interaction parameters are 
$U=6$ eV and $\lambda=0.95$.

\begin{thebibliography}{99}
\bibitem{den Hertog}
    B.C. den Hertog and M.P. Das, Solid State Commun. {\bf 98}, 7 (1996). Note that in this reference there are  small typographical
errors in equations (3) and (12). In equation (3) the square root terms should be in the denominator. In equation (12) the derivative should be w.r.t. $n_{i}$, not $d_{i}$.
None of the results are effected by these errors.
\end{thebibliography}
\end{document}